\documentclass{article}
\usepackage{graphicx}
\usepackage{subfigure}
\usepackage{cite}

\def\met{\ensuremath{E_{\mathrm{T}}^{\mathrm{miss}}}}

\title{Vector-like $B$ pair production at future $pp$ colliders}

\author{Erich W. Varnes}

\begin{document}

\maketitle

\section{Introduction}

Heavy vector-like quarks (VLQ) appear in many extensions to the SM that provide a solution to the hierarchy problem~\cite{Perelstein04,Contino07,Matsedonskyi13,Kaplan91,Contino07_2,Hewett89}.  Searches for such quarks are underway at the LHC, but these are currently not sensitive to VLQ masses greater than about 750 GeV.  Here the sensitivity of future accelerators to pair-produced vector-like $B$ quarks is studied.

\section{Samples and scenarios}

All samples used for these studies were processed with the {\sc DELPHES}~\cite{delphes} fast detector simulation, using the generic ``Snowmass detector'' parameters~\cite{snowmassdetector}.  The background samples were generated in bins of $H_T$, as described in~\cite{bkgsamples}.  The SM background sources considered were diboson and $t\bar{t}+W/Z$ production.  There are also backgrounds from detector effects leading to hadronic jets appearing as leptons or to the mis-measurement of the lepton charge.  The magnitude of these backgrounds will depend on the details of the detector and of the environment in which it operates; here we estimate the contribution by also considering $t\bar{t}$ events as a background.  Such events never contain same-sign isolated leptons pairs, but may appear to if there is a fake lepton or one with mis-measured charge.  Signal samples were generated with {\sc mad graph}~\cite{madgraph} v1.5.10, with $B$ masses of 1.0 and 1.5 TeV.  These events were then processed by {\sc pythia}8~\cite{pythia} to simulate decays and hadronization. The decays of the $B$ or $\bar{B}$ was forced to $Wt$, $Zb$, or $Hb$ such that each of the six possible $B\bar{B}$ final states ($WtWt$, $WtZb$, $WtHb$, $ZbZb$, $ZbHb$, or $HbHb$) were generated as separate samples.  There samples were then processed by {\sc DELPHES} with 0, 50, or 140 minimum-bias events overlaid on average to simulate the effects of pileup.  The cross sections for all samples used are shown in Table~\ref{tab:mcxsec}.

\begin{table}
\caption{\label{tab:mcxsec} 
Cross sections in pb for the MC samples used in this study.  The cross sections are NLO for all samples.  The symbol B refers to a vector boson, while $B$ refers to a vector-like $b$ quark.}
\begin{center}
\begin{tabular}{lcc}
Sample & $\sqrt{s}$ (TeV)  & Cross section (pb)  \\ \hline 
$B\bar{B}$, $m=1000$ GeV & 14 & 0.059  \\
 $B\bar{B}$, $m=1500$ GeV & 14 & 0.0030\\
 BB-4p-0-300 & 14 & 325 \\
 BB-4p-300-700  & 14 & 44.8 \\
 BB-4p-700-1300 & 14 & 5.4 \\
 BB-4p-1300-2100 & 14 &0.53 \\
 BB-4p-2100-100000 & 14 & 0.060\\
 ttB-4p-0-900 & 14 & 2.65 \\
 ttB-4p-900-1600 & 14 & 0.25 \\
 ttB-4p-1600-2500 & 14 & 0.024 \\
 ttB-4p-2500-100000 & 14 & 0.0021 \\
 tt-4p-0-600 & 14 & 664 \\
 tt-4p-600-1100 & 14  & 51 \\
 tt-4p-1100-1700 & 14 & 5.3 \\
 tt-4p-1700-2500 & 14 & 0.68 \\
 tt-4p-2500-100000 & 14 & 0.067 \\ \hline
 $B\bar{B}$, $m=1000$ GeV & 33 & 1.28  \\
 $B\bar{B}$, $m=1500$ GeV & 33 & 0.13\\
 BB-4p-0-400 & 33 & 792 \\
 BB-4p-400-1000 & 33 & 109 \\
 BB-4p-1000-2000 & 33 & 10.5 \\
 BB-4p-2000-3400 & 33 & 0.86 \\
 BB-4p-3400-100000 & 33 & 0.10 \\
 ttB-4p-0-1200 & 33 & 23.9 \\
 ttB-4p-1200-2200 & 33 & 1.58\\
 ttB-4p-2200-3600 & 33 & 0.16 \\
 ttB-4p-3600-100000 & 33 & 0.017 \\
 tt-4p-0-600 & 33 & 3955 \\
 tt-4p-600-1200 & 33  & 582 \\
 tt-4p-1200-2000 & 33 & 71 \\
 tt-4p-2000-3200 & 33 & 8.8 \\
 tt-4p-3200-4800 & 33 & 0.83 \\
 tt-4p-4800-100000 & 33 & 0.032 \\
\hline
\end{tabular} 
\end{center}
\end{table}
 
The following $pp$ collider  scenarios are considered:
\begin{itemize}
\item  $\sqrt{s} = 14$ TeV, with an average of 50 pileup events and integrated luminosity of 300 fb$^{-1}$.
\item   $\sqrt{s} = 14$ TeV, with an average of 140 pileup events and integrated luminosity of 3000 fb$^{-1}$.
\item   $\sqrt{s} = 33$ TeV, with an average of 50 pileup events and integrated luminosity of 300 fb$^{-1}$ (in progress).
\end{itemize}
 
\section{Event selection}

In the same-sign dilepton channel, events are required to have exactly one pair of leptons with the same electric charge, at least two jets (one of which must satisfy a loose $b$ ID requirement).  To exploit the full range of $B$ decay scenarios, the branching fractions are scanned in steps of 0.01.  For each assumed set of branching fractions, the criteria imposed on $H_T$, $\met$, and  the number of $b$-tagged jets are set such that the quantity
$$
S \equiv {N_s \over \sqrt{N_s^2+N_b^2 + \sigma_{\rm syst}^2}}
$$
is maximized, where $N_s$ and $N_b$ are the expected signal and background yields, respectively, and $\sigma_{\rm syst}$ is the  systematic uncertainty.  For these studies, $\sigma_{\rm syst}$ is parameterized as a fraction of $N_b$.

\section{Results}

In lieu of a complete calculation of confidence intervals, $S$ is used to estimate the sensitivity of future accelerators to the vector-like $B$.  If $S$ is greater than 2 for a given $B$ mass and set of branching fractions, that hypothesis is considered to be one that may be excluded; if $S$ is greater than 3, then the accelerator has the potential to find evidence for the hypotheses, and if $S$ is greater than 5, the accelerator is considered to have discovery potential.

\begin{figure}[htp]
\subfigure[]{\includegraphics[width=7 cm]{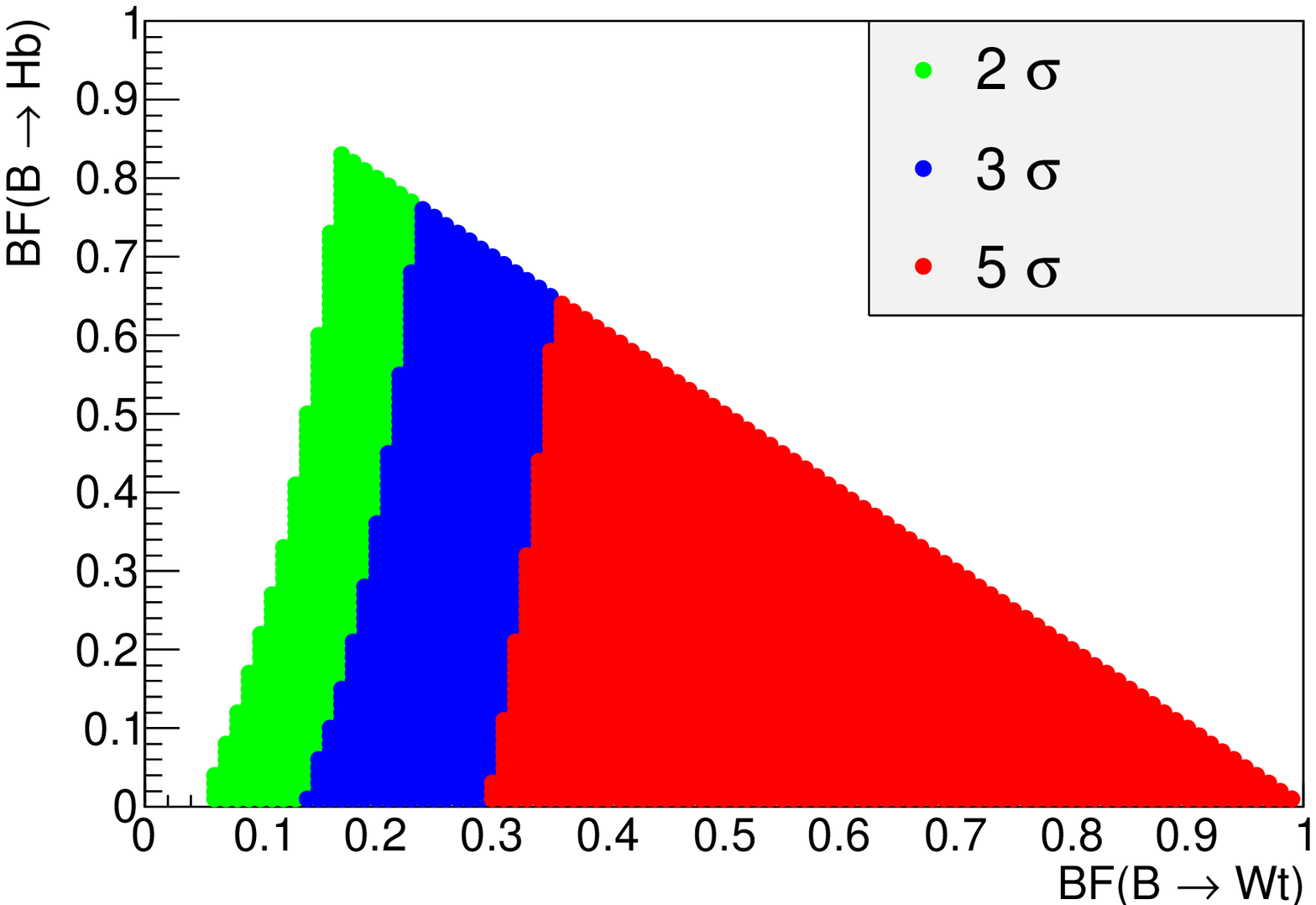}}
\subfigure[]{\includegraphics[width=7 cm]{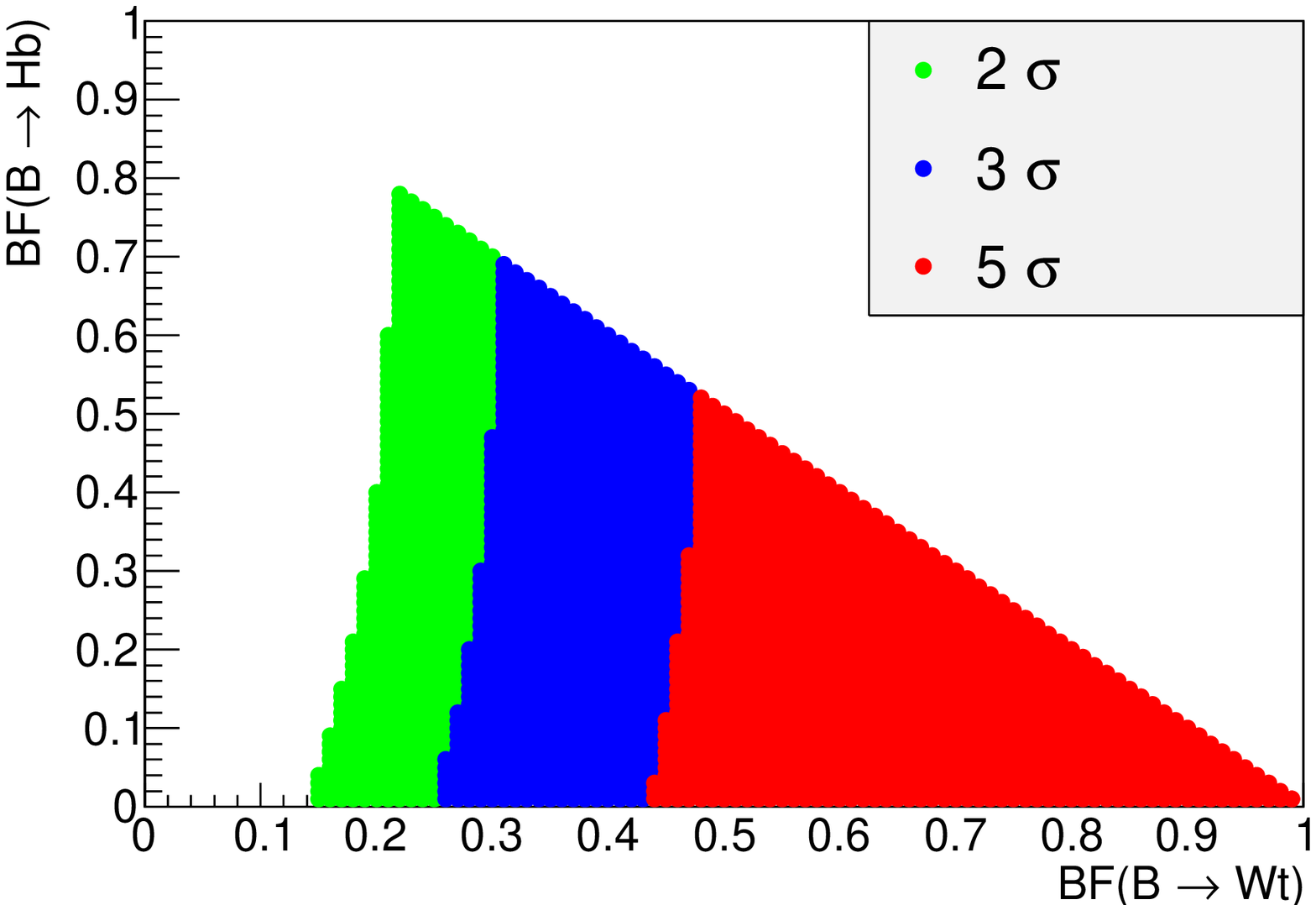}}
\subfigure[]{\includegraphics[width=7 cm]{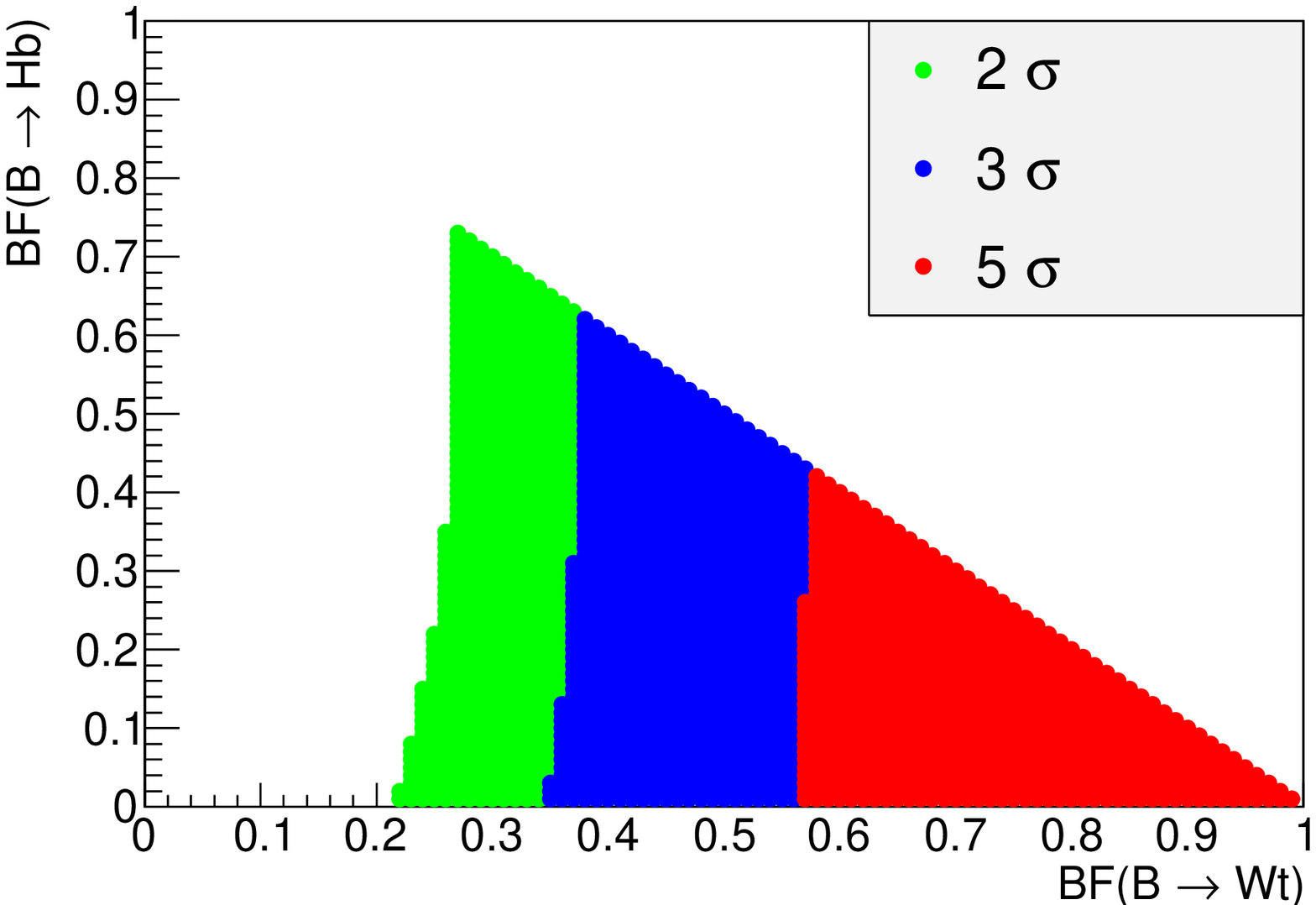}}
\caption{\label{fig:m1000_14TeV_300} Expected sensitivity to pair production of 1000 GeV $B$ quarks at a 14 TeV LHC, assuming an average of 50 interactions per bunch crossing and an integrated luminosity of 300 fb$^{-1}$.  The assumed systematic uncertainty, expressed as a fraction of the background yield, is 0 for (a), 20\% for (b), and 40\% for (c).}
\end{figure}

\begin{figure}[htp]
\subfigure[]{\includegraphics[width=7 cm]{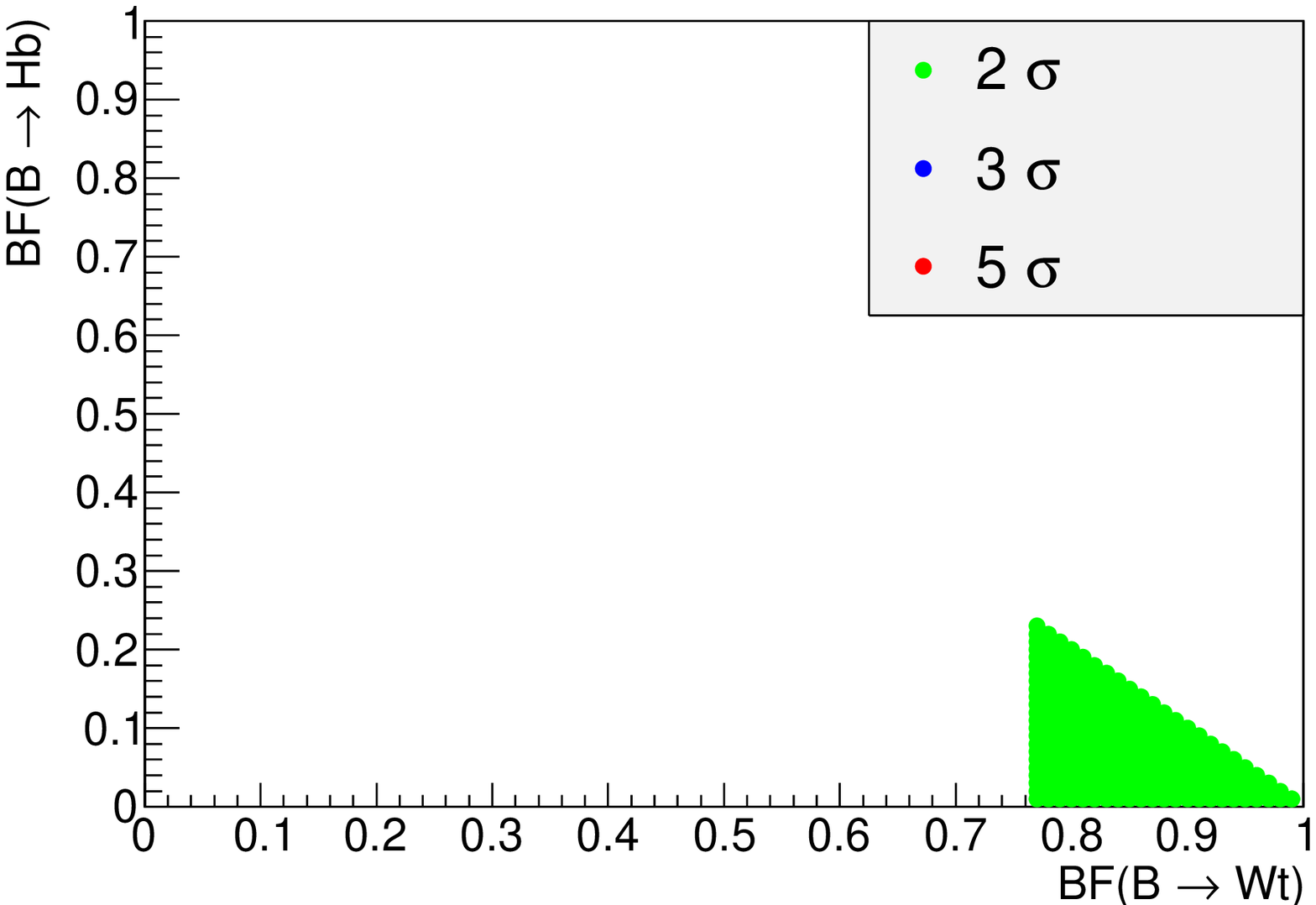}}
\subfigure[]{\includegraphics[width=7 cm]{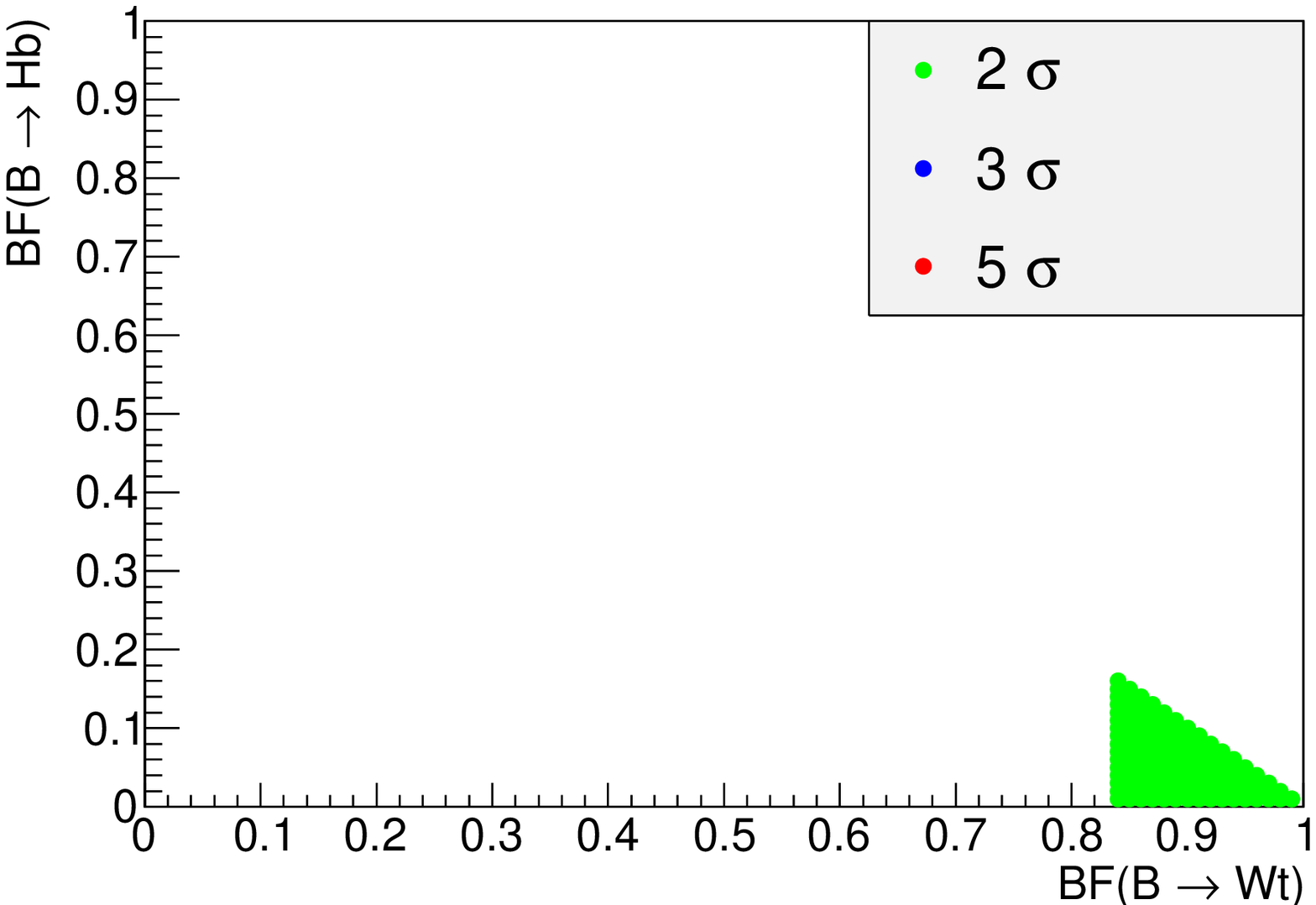}}
\subfigure[]{\includegraphics[width=7 cm]{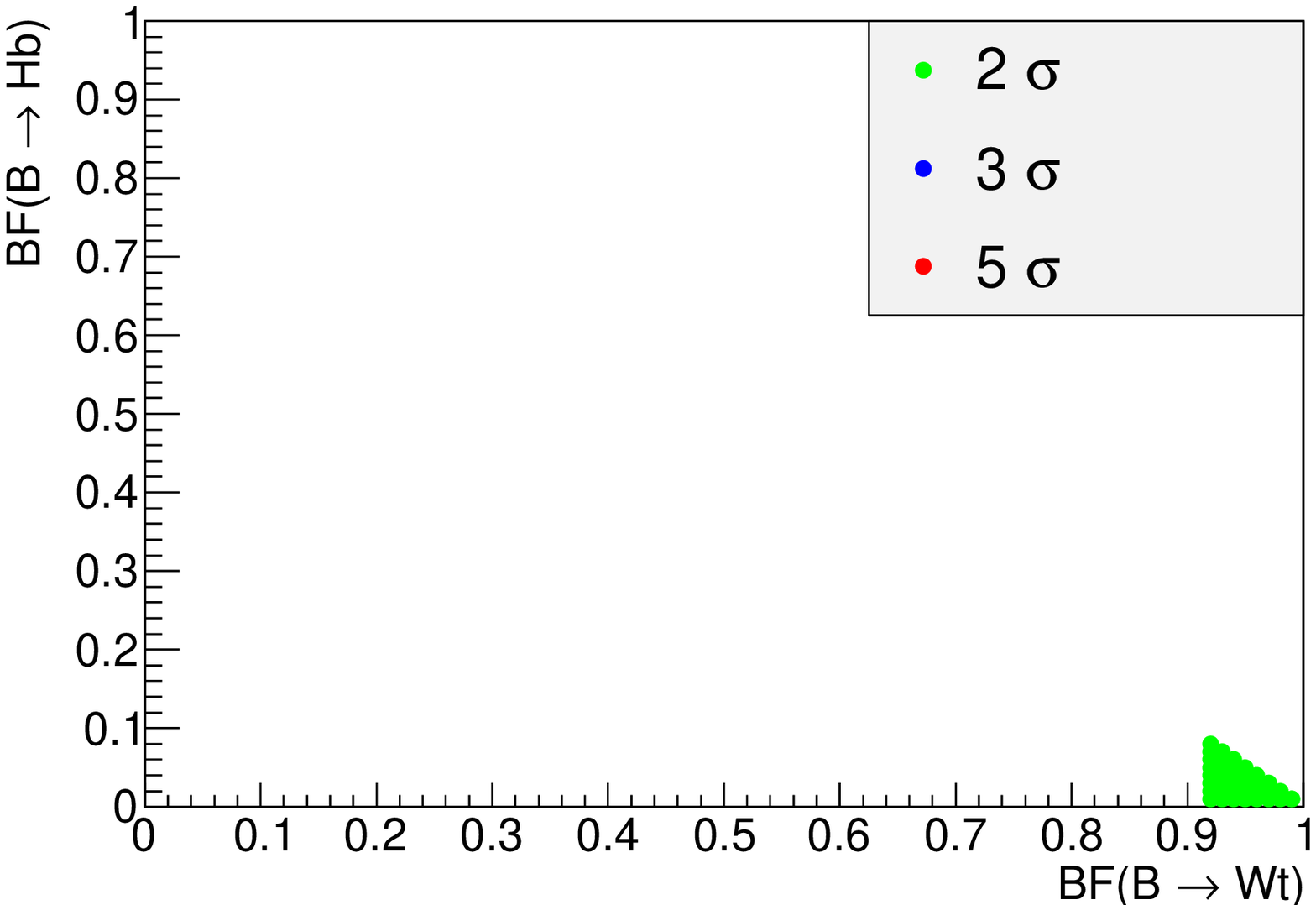}}
\caption{\label{fig:m1500_14TeV_300} Expected sensitivity to pair production of 1500 GeV $B$ quarks at a 14 TeV LHC, assuming an average of 50 interactions per bunch crossing and an integrated luminosity of 300 fb$^{-1}$.  The assumed systematic uncertainty, expressed as a fraction of the background yield, is 0 for (a), 20\% for (b), and 40\% for (c).}
\end{figure}

\begin{figure}[htp]
\subfigure[]{\includegraphics[width=7 cm]{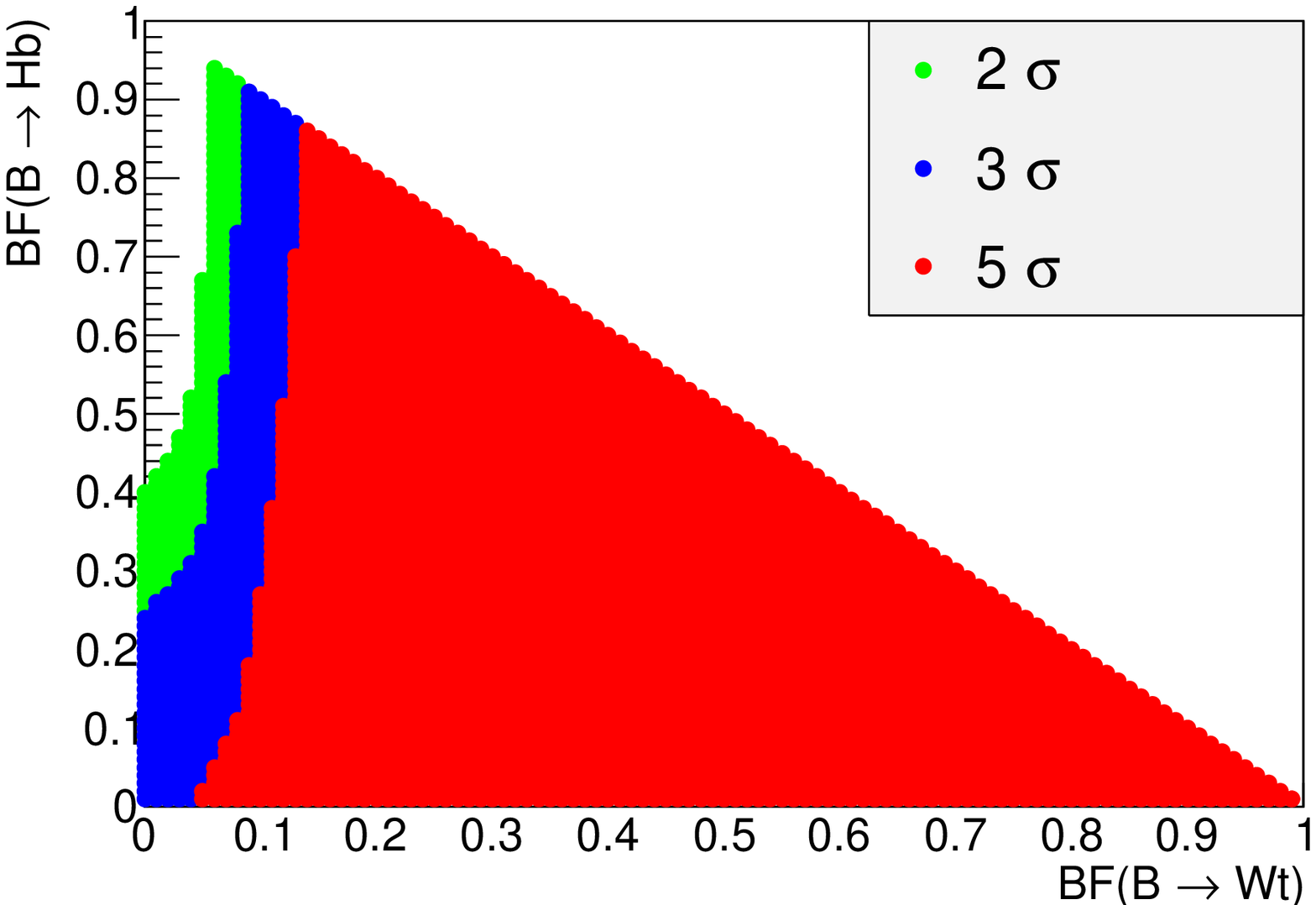}}
\subfigure[]{\includegraphics[width=7 cm]{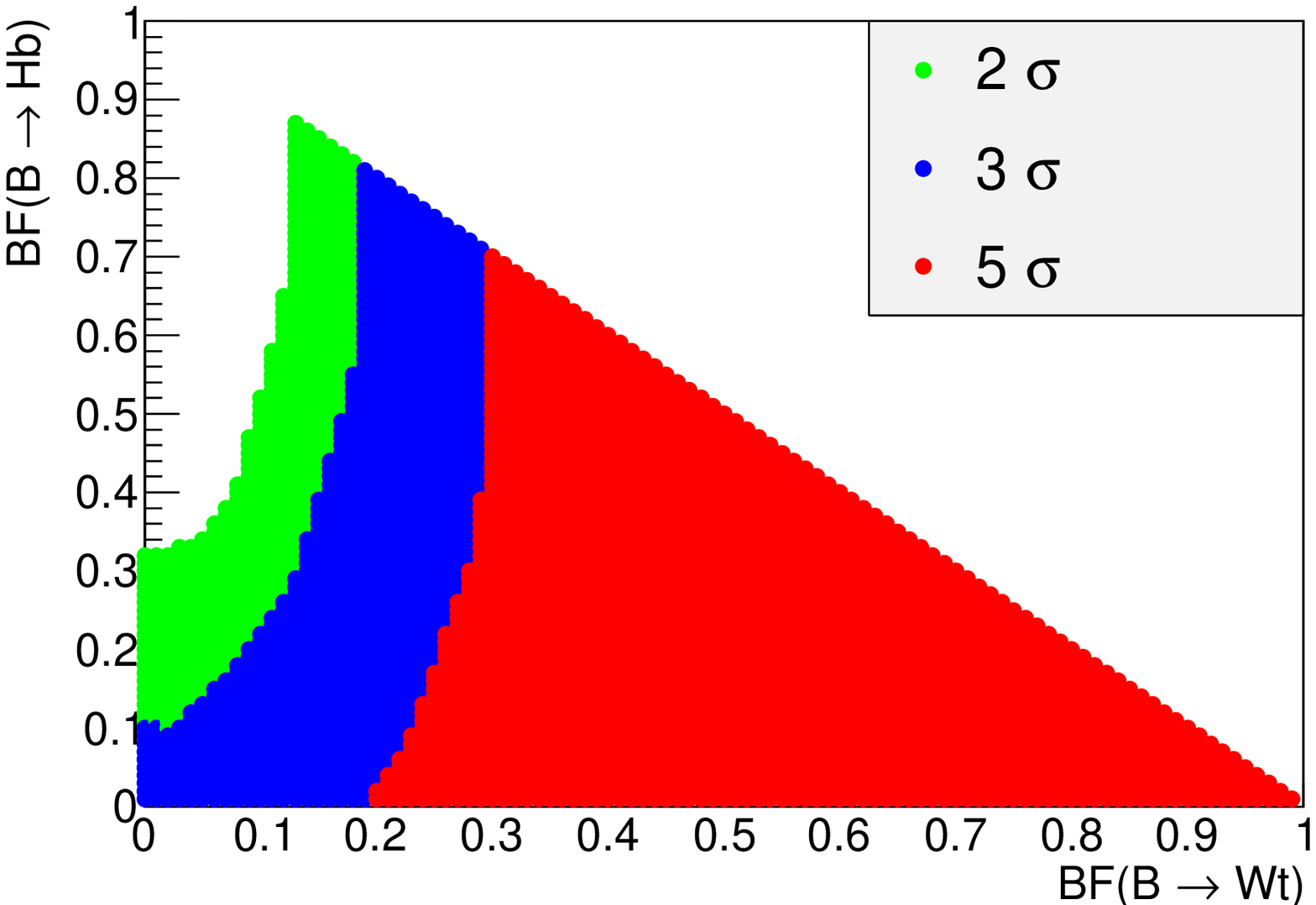}}
\subfigure[]{\includegraphics[width=7 cm]{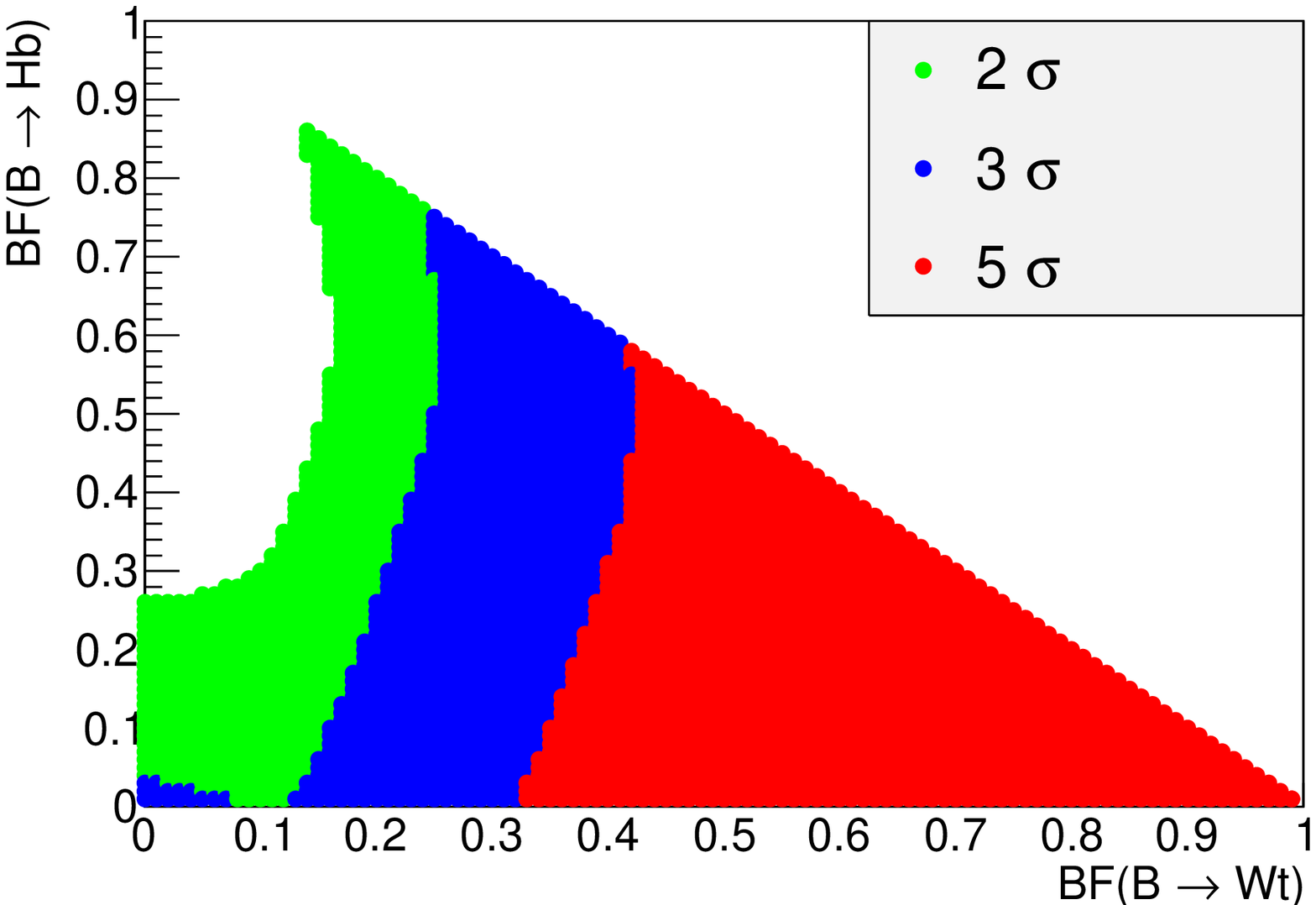}}
\caption{\label{fig:m1000_14TeV_3000} Expected sensitivity to pair production of 1000 GeV $B$ quarks at a 14 TeV LHC, assuming an average of 140 interactions per bunch crossing and an integrated luminosity of 3000 fb$^{-1}$.  The assumed systematic uncertainty, expressed as a fraction of the background yield, is 0 for (a), 20\% for (b), and 40\% for (c).}
\end{figure}

\begin{figure}[htp]
\subfigure[]{\includegraphics[width=7 cm]{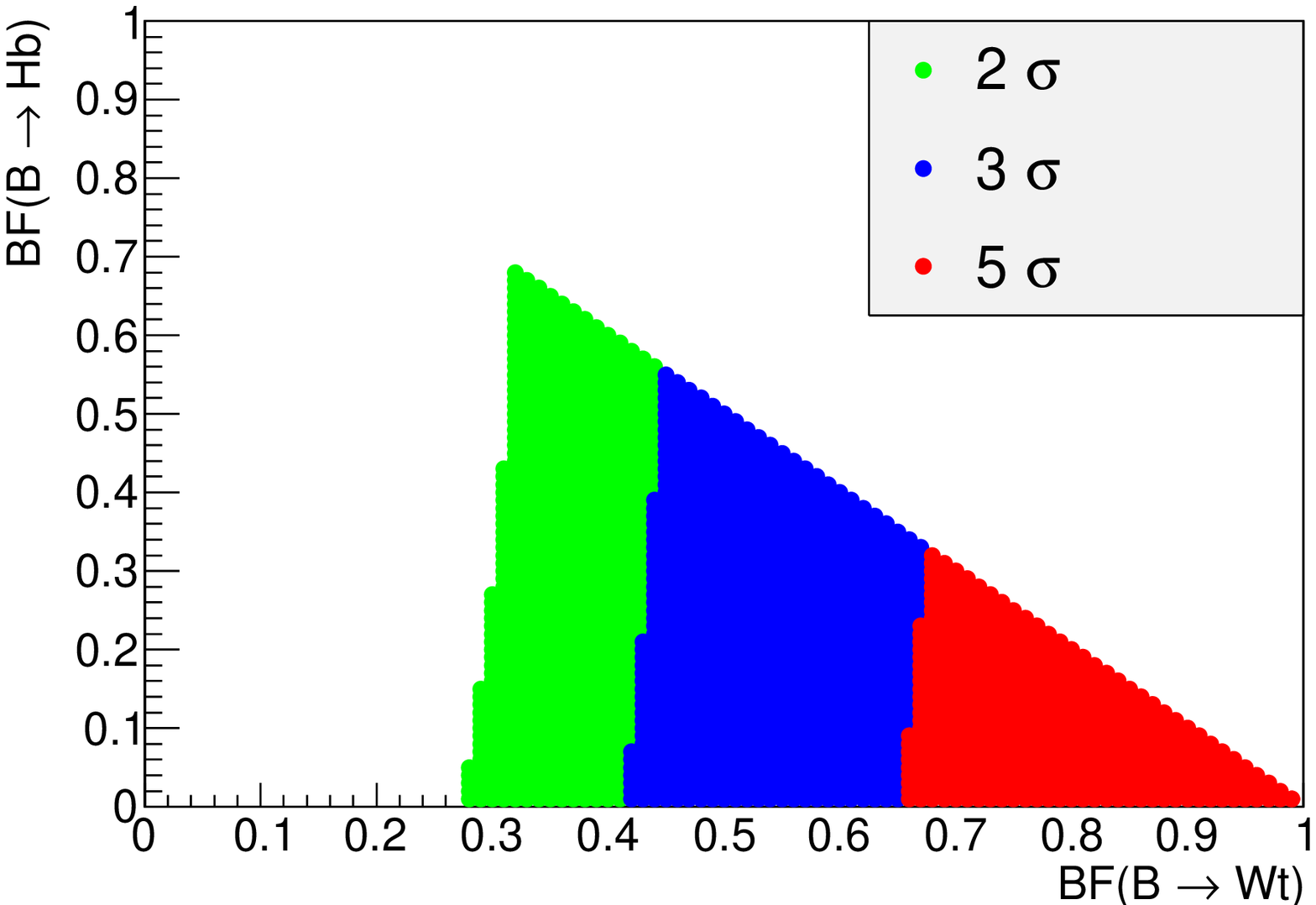}}
\subfigure[]{\includegraphics[width=7 cm]{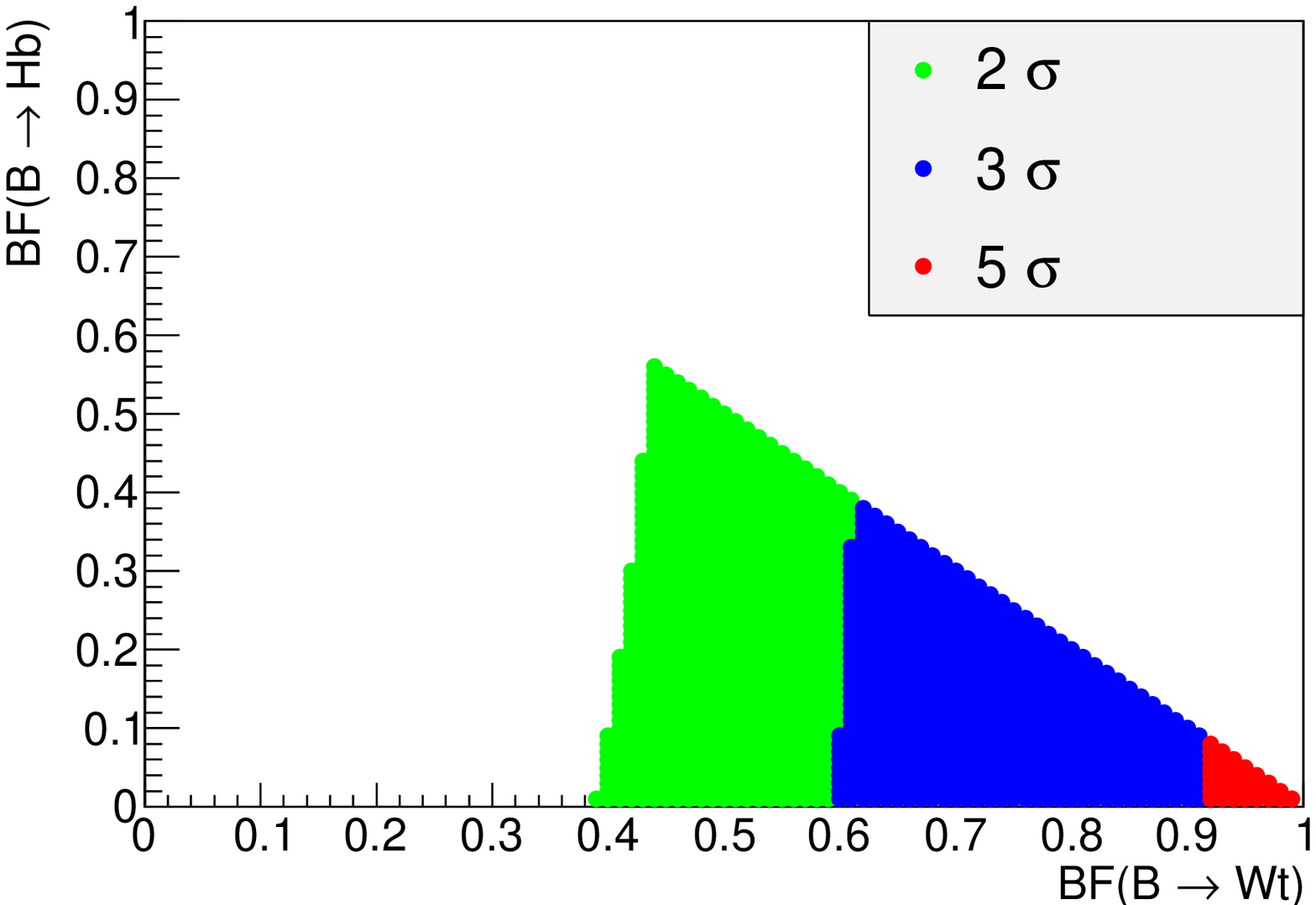}}
\subfigure[]{\includegraphics[width=7 cm]{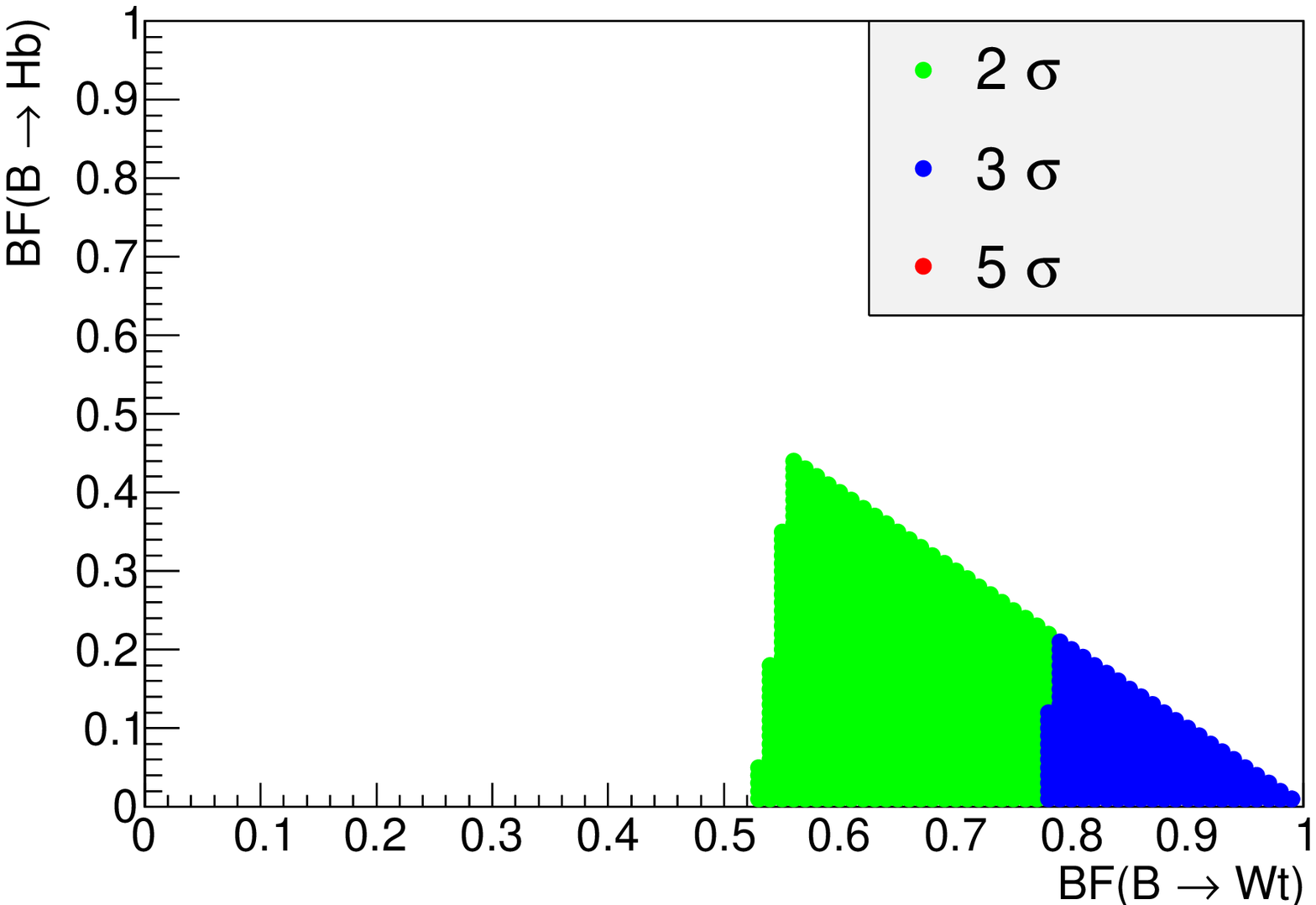}}
\caption{\label{fig:m1500_14TeV_3000} Expected sensitivity to pair production of 1500 GeV $B$ quarks at a 14 TeV LHC, assuming an average of 140 interactions per bunch crossing and an integrated luminosity of 3000 fb$^{-1}$.  The assumed systematic uncertainty, expressed as a fraction of the background yield, is 0 for (a), 20\% for (b), and 40\% for (c).}
\end{figure}

Results are shown in Figs.~\ref{fig:m1000_14TeV_300}-\ref{fig:m1500_14TeV_3000}.  These show that a 14 TeV LHC with 300 fb$^{-1}$ would be quite sensitive to a 1000 GeV $B$ quark, having discovery potential for some branching fractions and exclusion potential for most branching fractions even under the most pessimistic assumption concerning systematic uncertainties.  For a $B$ quark with mass 1500 GeV, the 300 fb$^{-1}$ dataset would provide little sensitivity, while a high-luminosity LHC that collection 3000 fb$^{-1}$ would be able to provide exclusion potential for a wide range of branching fractions.

\begin{figure}[htp]
\subfigure[]{\includegraphics[width=7 cm]{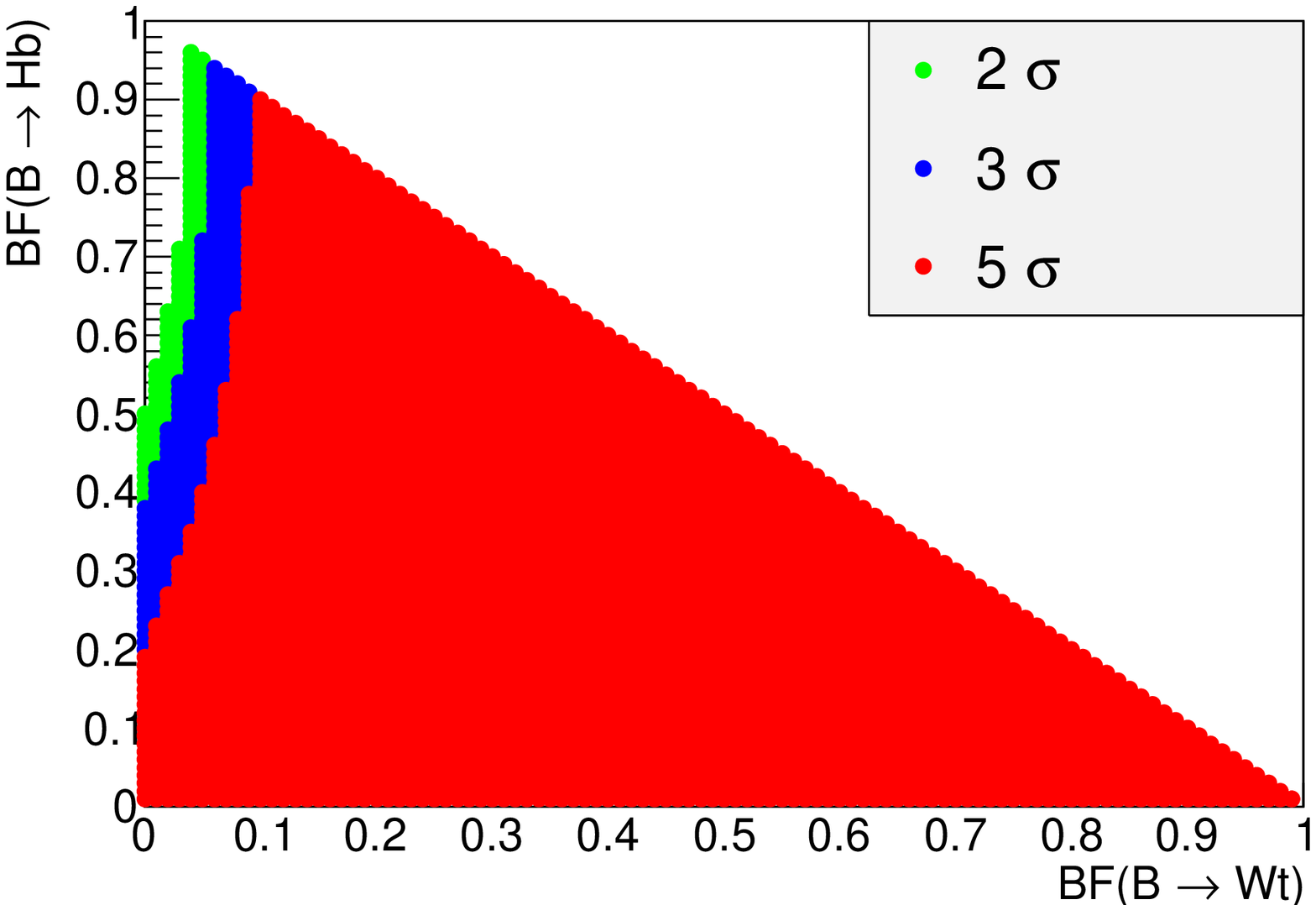}}
\subfigure[]{\includegraphics[width=7 cm]{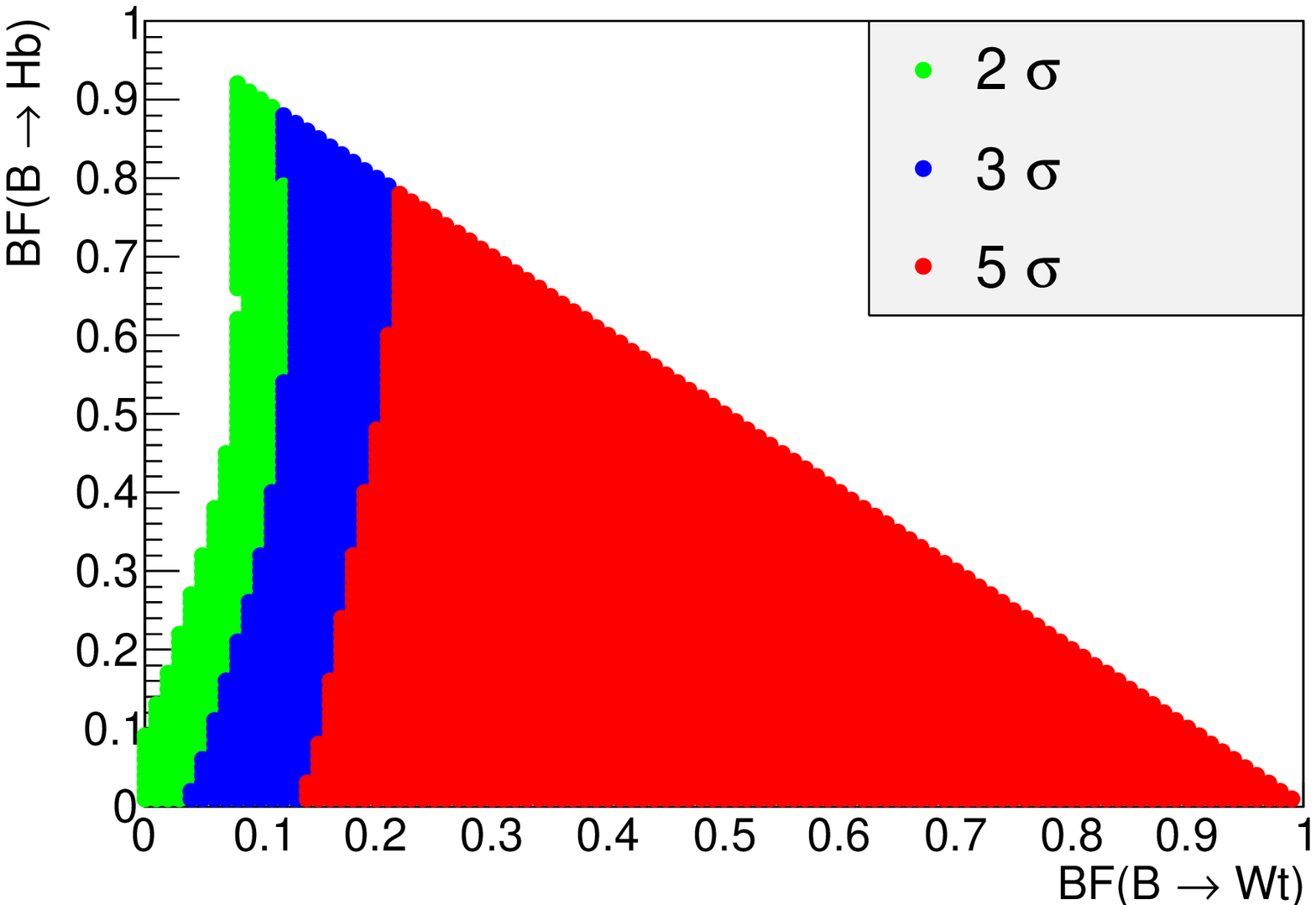}}
\subfigure[]{\includegraphics[width=7 cm]{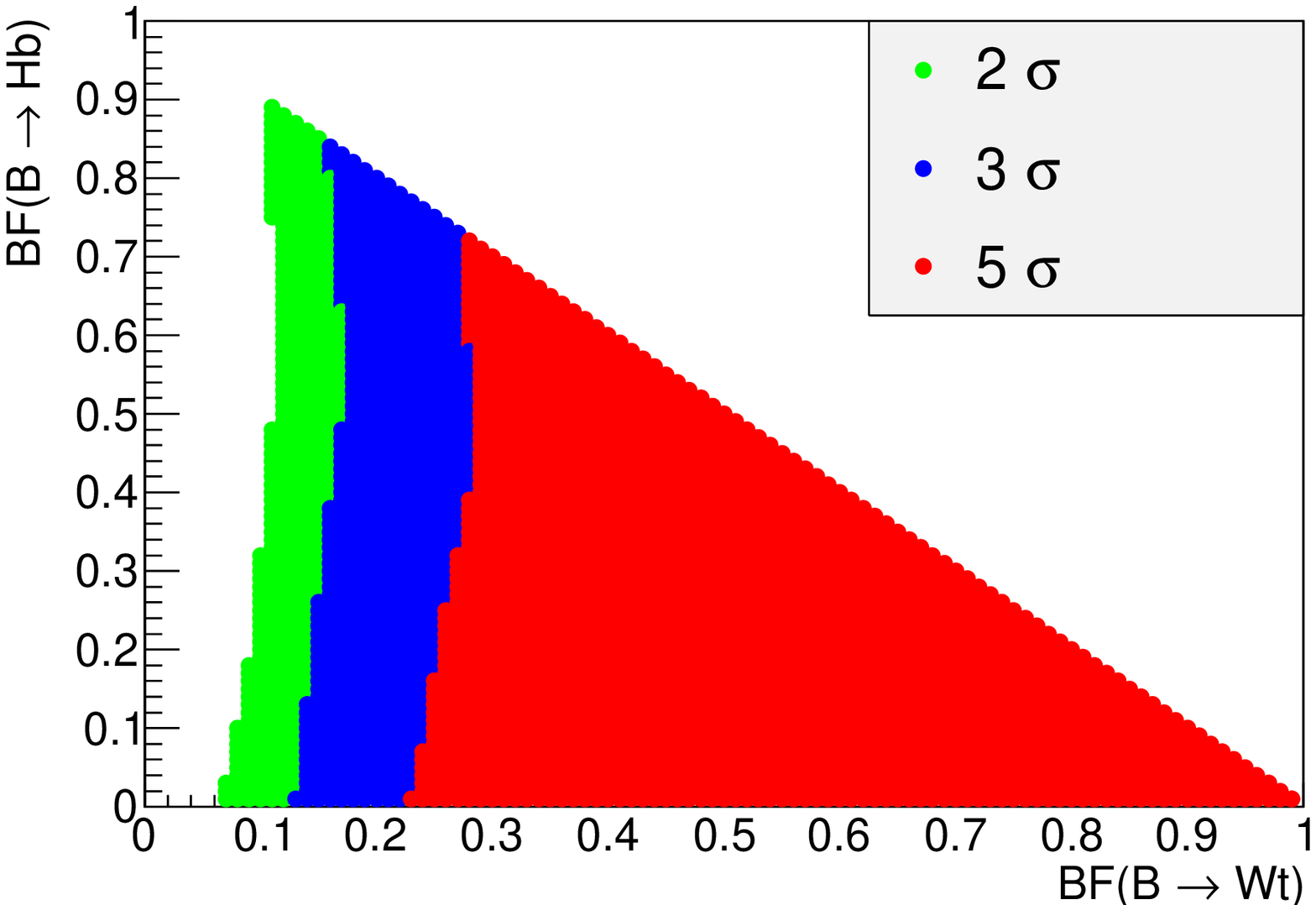}}
\caption{\label{fig:m1000_33TeV_300} Expected sensitivity to pair production of 1000 GeV $B$ quarks at a 33 TeV LHC, assuming an average of 50 interactions per bunch crossing and an integrated luminosity of 300 fb$^{-1}$.  The assumed systematic uncertainty, expressed as a fraction of the background yield, is 0 for (a), 20\% for (b), and 40\% for (c).}
\end{figure}

\begin{figure}[htp]
\subfigure[]{\includegraphics[width=7 cm]{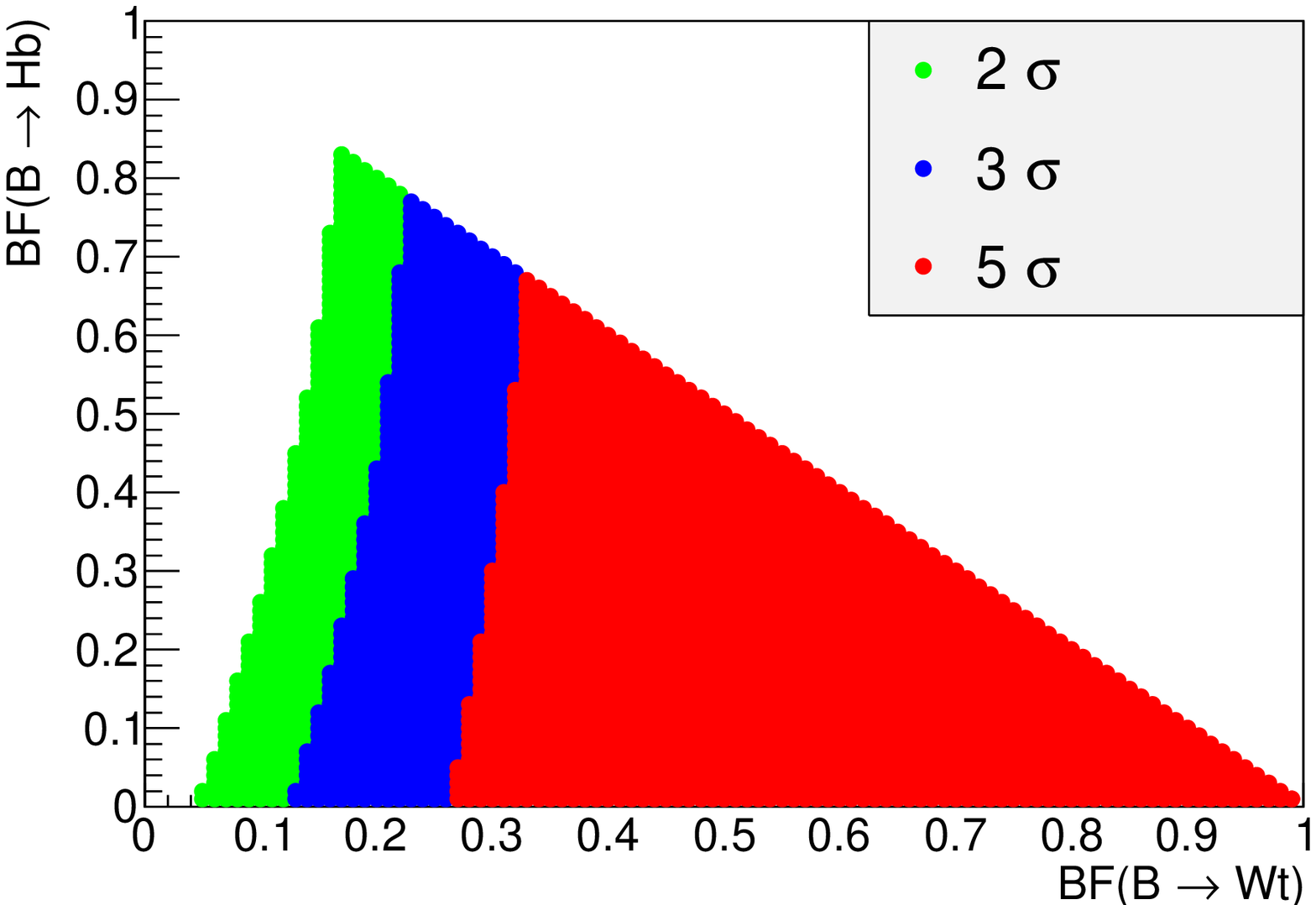}}
\subfigure[]{\includegraphics[width=7 cm]{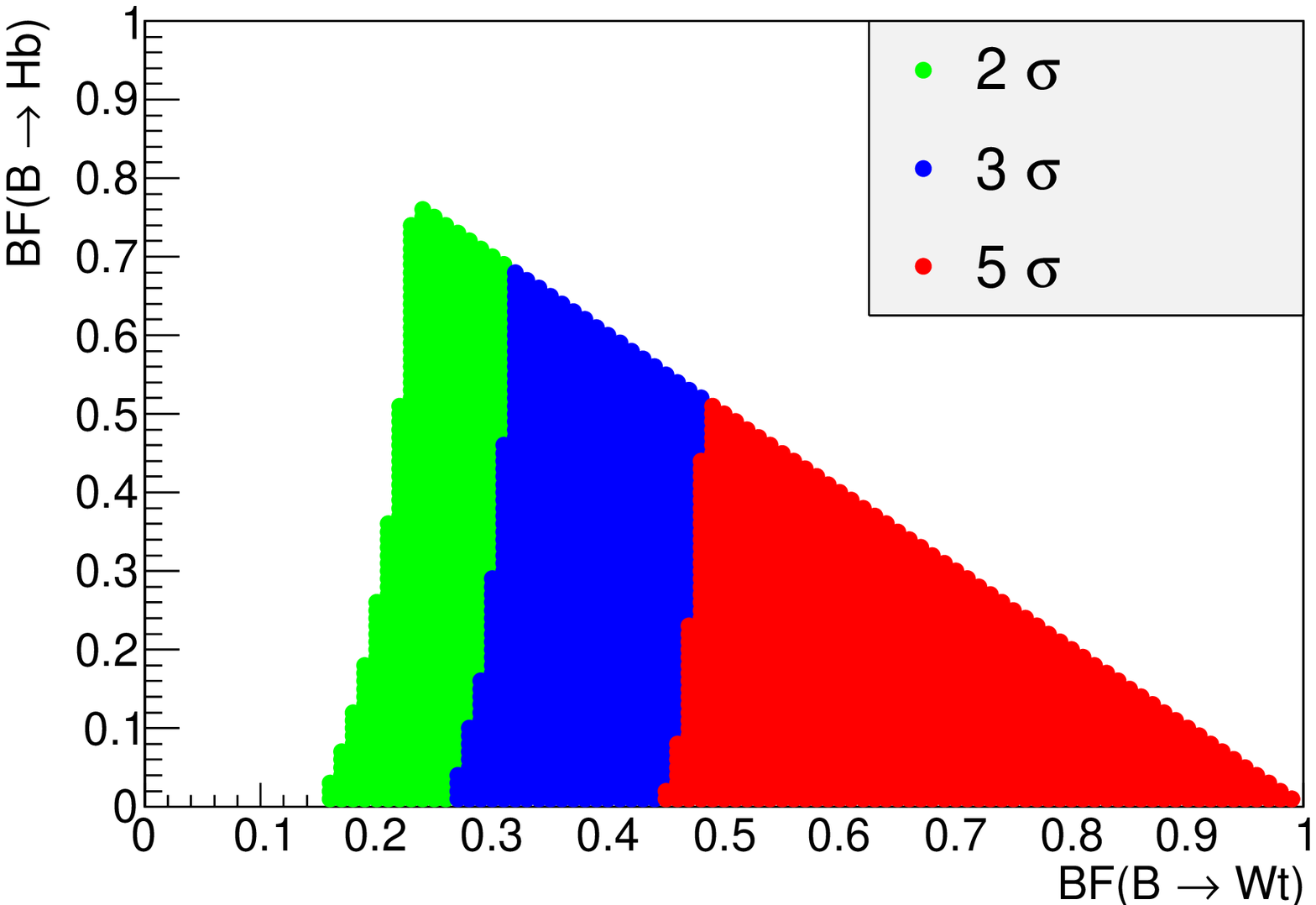}}
\subfigure[]{\includegraphics[width=7 cm]{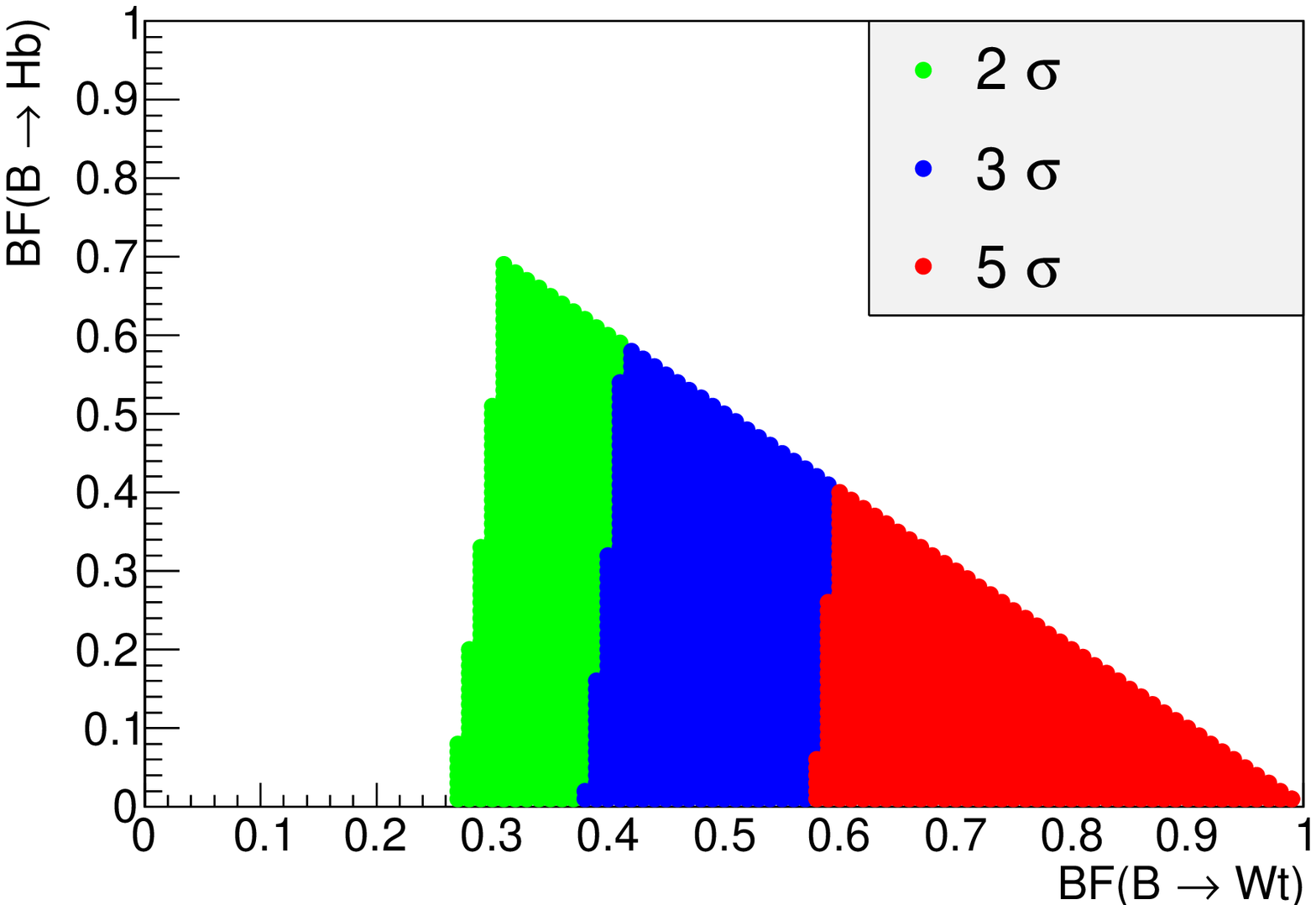}}
\caption{\label{fig:m1500_33TeV_300} Expected sensitivity to pair production of 1500 GeV $B$ quarks at a 33 TeV LHC, assuming an average of 50 interactions per bunch crossing and an integrated luminosity of 300 fb$^{-1}$.  The assumed systematic uncertainty, expressed as a fraction of the background yield, is 0 for (a), 20\% for (b), and 40\% for (c).}
\end{figure}

For the 33 TeV LHC, results are shown in Figs.~\ref{fig:m1000_33TeV_300} and \ref{fig:m1500_33TeV_300}.  These clearly demonstrate the advantage of greater energy for this search, substantially expanding the range of branching ratios to which the search would be sensitive for a 1500 GeV $B$ quark.

In addition to the coverage for arbitrary branching ratios presented in Figs.~\ref{fig:m1000_14TeV_300}-\ref{fig:m1500_33TeV_300}, it is useful to consider a particular benchmark set of branching fractions.  One that is motivated by naturalness is  BF$(B \rightarrow Wt) = 0.5$,  BF$(B \rightarrow Zb) =$  BF$(B \rightarrow Hb) = 0.25$.
The optimized signal and background yields for this set of branching ratios are given in Tables~\ref{tab:sigbkg_m1000_14TeV}-\ref{tab:sigbkg_m1500_33TeV}.  One can also compute the upper limit on the $B$ quark mass for a given signal significance using these assumed branching fractions.  The results, for the scenario where the systematic uncertainties are 20\% of the background yield, are given in Table~\ref{tab:mass_limits}.

\begin{table}
\caption{\label{tab:sigbkg_m1000_14TeV} 
Expected signal and background yields for the optimized selection at $\sqrt{s} = 14$ TeV, assuming that the mass of the $B$ is 1000 GeV,  BF$(B \rightarrow Wt)=0.5$, and BF$(B \rightarrow Zb) =$  BF$(B \rightarrow Hb) = 0.25$. The average number of interactions per bunch crossing is represented by $\mu$.}
\begin{center}
\begin{tabular}{ccccccc}
\hline
 Scenario & \multicolumn{3}{c}{${\cal L} = 300$ fb$^{-1}$, $\mu = 50$} & \multicolumn{3}{c}{${\cal L} = 3000$ fb$^{-1}$, $\mu = 140$}  \\  \hline
Systematics & 0 & 20\% & 40\% & 0 & 20\% & 40\% \\ \hline
 Diboson & 94 & 4.3 & 2.0 & 177 & 28.4 & 9.2 \\
 $t\bar{t}+W/Z$ & 183 & 30.6 & 11.1 & 846 & 78.3 & 17.9  \\
 $t\bar{t}$ & 48 & 4.4 & 1.8 & 420 & 41.1 & 10.0  \\ \hline
 Total bkg & 325 & 39.3 & 15.0 & 1442 & 148 & 37.0  \\ \hline
 Signal & 181 & 71.5 & 15.0 & 1202 & 355 & 124  \\ \hline
\hline
\end{tabular} 
\end{center}
\end{table}

\begin{table}
\caption{\label{tab:sigbkg_m1500_14TeV} 
Expected signal and background yields for the optimized selection at $\sqrt{s} = 14$ TeV, assuming that the mass of the $B$ is 1500 GeV,  BF$(B \rightarrow Wt)=0.5$, and BF$(B \rightarrow Zb) =$  BF$(B \rightarrow Hb) = 0.25$. The average number of interactions per bunch crossing is represented by $\mu$.}
\begin{center}
\begin{tabular}{ccccccc}
\hline
 Scenario & \multicolumn{3}{c}{${\cal L} = 300$ fb$^{-1}$, $\mu = 50$} & \multicolumn{3}{c}{${\cal L} = 3000$ fb$^{-1}$, $\mu = 140$}  \\  \hline
Systematics & 0 & 20\% & 40\% & 0 & 20\% & 40\% \\ \hline
 Diboson & 1.5 & 0.8 & 0.8 & 15.7 & 6.7 & 0.8  \\
 $t\bar{t}+W/Z$ & 7.0 & 2.5 & 2.5 & 37.9 & 9.0 & 2.8 \\
 $t\bar{t}$ & 1.3 & 0.4 & 0.4 & 20.3 & 4.8 & 1.9 \\ \hline
 Total bkg & 9.8 & 3.7 & 3.7 & 73.9 & 20.5 & 5.4  \\ \hline
 Signal & 4.36 & 2.9 & 2.9 & 37.3 & 18.0 & 7.7 \\ \hline
\hline
\end{tabular} 
\end{center}
\end{table}

\begin{table}
\caption{\label{tab:sigbkg_m1000_33TeV} 
Expected signal and background yields for the optimized selection at $\sqrt{s} = 33$ TeV, assuming that the mass of the $B$ is 1000 GeV,  BF$(B \rightarrow Wt)=0.5$, and BF$(B \rightarrow Zb) =$  BF$(B \rightarrow Hb) = 0.25$. The average number of interactions per bunch crossing is represented by $\mu$.}
\begin{center}
\begin{tabular}{cccc}
\hline
 Scenario & \multicolumn{3}{c}{${\cal L} = 300$ fb$^{-1}$, $\mu = 50$}  \\  \hline
Systematics & 0 & 20\% & 40\%  \\ \hline
 Diboson & 547 & 8.5 & 7.0  \\
 $t\bar{t}+W/Z$ & 1656 & 62.3 & 54.0   \\
 $t\bar{t}$ & 1628 & 20.8 & 18.2  \\ \hline
 Total bkg & 3302 & 91.7 & 79.2  \\ \hline
 Signal & 3040 & 390 & 356.4  \\ \hline
\hline
\end{tabular} 
\end{center}
\end{table}

\begin{table}
\caption{\label{tab:sigbkg_m1500_33TeV} 
Expected signal and background yields for the optimized selection at $\sqrt{s} = 33$ TeV, assuming that the mass of the $B$ is 1500 GeV,  BF$(B \rightarrow Wt)=0.5$, and BF$(B \rightarrow Zb) =$  BF$(B \rightarrow Hb) = 0.25$. The average number of interactions per bunch crossing is represented by $\mu$.}
\begin{center}
\begin{tabular}{cccc}
\hline
 Scenario & \multicolumn{3}{c}{${\cal L} = 300$ fb$^{-1}$, $\mu = 50$}   \\  \hline
Systematics & 0 & 20\% & 40\% \\ \hline
 Diboson & 102 & 3.1 & 3.1  \\
 $t\bar{t}+W/Z$ & 236 & 17.2 & 17.2  \\
 $t\bar{t}$ & 118 & 5.8 & 5.8 \\ \hline
 Total bkg & 456 & 26.2 & 26.2   \\ \hline
 Signal & 245 & 56.0 & 56.0 \\ \hline
\hline
\end{tabular} 
\end{center}
\end{table}

\begin{table}
\caption{\label{tab:mass_limits} 
Expected upper limits on the $B$ quark mass (in GeV) for a given statistical significance, assuming that BF$(B \rightarrow Wt)=0.5$ and BF$(B \rightarrow Zb) =$  BF$(B \rightarrow Hb) = 0.25$.}
\begin{center}
\begin{tabular}{ccccc}
\hline
 $\sqrt{s}$ (TeV) & ${\cal L}$ (fb$^{-1}$) & 2$\sigma$ & 3$\sigma$ & 5$\sigma$\\  \hline
14   &  300 & 1330 & 1210 & 1080 \\ 
	& 3000 & $> 1500$ & 1490 & 1330 \\
 33   & 300 & $> 1500$  & $> 1500$  & $> 1500$ \\ \hline
 \end{tabular} 
\end{center}
\end{table}

\section{Discussion}

It should be noted that there are several limitations to this study.  Among these are that some physics effects that could reduce the efficiency for signal or increase background are not available within the {\sc delphes} fast simulation, and there has been no attempt to rigorously estimate the systematic uncertainties for future accelerators.  On the other hand, the analysis presented uses a simple square-cut event selection, and bases the sensitivity estimate on the counts of signal and background events expected to pass the selection. When the actual data from future accelerators is available, more sophisticated analysis tools, such as some type of multi-variate analysis, will surely be used.   This will not only provide better signal/background discrimination than estimated here, but also allow direct measurement of the background normalization by counting events with low MVA values.  

It should also be noted that the observation of an excess in the same-sign lepton plus $b$ jet and \met final state could not be directly attributed to $B$ pair production, as there are other exotic phenomena that could cause such an excess.  Narrowing the possibilities will require looking for excesses in other final states (lepton plus jets, selections with reconstructed $Z$ bosons, etc.) and will therefore require more data than the initial observation.

\end{document}